\documentclass[twocolumn]{aastex63}

\usepackage{color}
\usepackage{amsmath}

\usepackage [english]{babel}
\usepackage [autostyle, english = american]{csquotes}
\MakeOuterQuote{"}

\makeatother

\shorttitle{VLBA Observations of Quasars with Offset Broad Lines}
\shortauthors{Breiding et al.}
\begin{document}
\setlength\abovedisplayskip{3pt}
\setlength\belowdisplayskip{3pt}
\title{The Search for Binary Supermassive Black Holes Amongst Quasars with Offset Broad Lines Using the Very Long Baseline Array}
\correspondingauthor{Peter Breiding}
\email{pbreiding@gmail.com}

\author[0000-0003-1317-8847]{Peter Breiding}
\affiliation{Department of Physics and Astronomy, West Virginia University, P.O. Box 6315, Morgantown, WV 26506, USA}
\affiliation{Center for Gravitational Waves and Cosmology, West Virginia University, Chestnut Ridge Research Building, Morgantown, WV 26505, USA}

\author[0000-0003-4052-7838]{Sarah Burke-Spolaor}
\affiliation{Department of Physics and Astronomy, West Virginia University, P.O. Box 6315, Morgantown, WV 26506, USA}
\affiliation{Center for Gravitational Waves and Cosmology, West Virginia University, Chestnut Ridge Research Building, Morgantown, WV 26505, USA}
\affiliation{Canadian Institute for Advanced Research, CIFAR Azrieli Global Scholar, MaRS Centre West Tower, 661 University Ave. Suite 505, Toronto ON M5G 1M1, Canada}

\author[0000-0002-3719-940X]{Michael Eracleous}
\affiliation{Department of Astronomy \& Astrophysics and Institute for Gravitation and the Cosmos, The Pennsylvania State University, 525 Davey Lab, University Park, PA 16802, USA}

\author[0000-0002-7835-7814]{Tamara Bogdanovi\'c}
\affiliation{School of Physics and Center for Relativistic Astrophysics, 837 State St NW, Georgia Institute of Technology, Atlanta, GA 30332, USA}

\author{T. Joseph W. Lazio}
\affiliation{Jet Propulsion Laboratory, California Institute of Technology, 4800 Oak Grove Dr, Pasadena, CA 91109, USA}

\author[0000-0001-8557-2822]{Jessie Runnoe}
\affiliation{Department of Physics and Astronomy, Vanderbilt University, 6301 Stevenson Circle, Nashville, TN 37235
	}

\author[0000-0002-8187-1144]{Steinn Sigurdsson}
\affiliation{Department of Astronomy \& Astrophysics and Institute for Gravitation and the Cosmos, The Pennsylvania State University, 525 Davey Lab, University Park, PA 16802, USA}

\begin{abstract}
In several previous studies, quasars exhibiting broad emission lines with $\gtrsim$1000~$\mathrm{km\ s^{-1}}$ velocity offsets with respect to the host galaxy rest frame have been discovered.  One leading hypothesis for the origin of these velocity-offset broad lines is the dynamics of a binary supermassive black hole (SMBH).  
We present high-resolution radio imaging of 34 quasars showing these velocity-offset broad lines with the Very Long Baseline Array (VLBA), aimed at finding evidence for the putative binary SMBHs (such as dual radio cores), and testing the competing physical models.  
We detect exactly half of the target sample from our VLBA imaging, after implementing a 5$\sigma$ detection limit.  While we do not resolve double radio sources in any of the targets, we obtain limits on the instantaneous projected separations of a radio-emitting binary for all of the detected sources under the assumption that a binary still exists within our VLBA angular resolution limits.  We also assess the likelihood that a radio-emitting companion SMBH exists outside of our angular resolution limits, but its radio luminosity is too weak to produce a detectable signal in the VLBA data.  Additionally, we compare the precise sky positions afforded by these data to optical positions from both the SDSS and \textit{Gaia}~DR2 source catalogs. We find projected radio/optical separations on the order of $\sim$10~pc for three quasars. Finally, we explore how future multi-wavelength campaigns with optical, radio, and X-ray observatories can help discriminate further between the competing physical models.  
\end{abstract}


\section{Introduction} \label{sec:intro}
\subsection{Binary SMBH Formation \& Evolution}
There is strong evidence that most massive galaxies host a supermassive black hole (SMBH, M$_{\mathrm{BH}}\gtrsim 10^{5}$M$_{\odot}$) at their  centers, and the mass of this black hole is tightly correlated with various host galaxy properties \citep[e.g.,][]{magorrian+98}.  SMBHs are believed to grow through two main channels: the accretion of surrounding gas and the hierarchical merging of other SMBHs via galaxy mergers \citep{kormendy13}.   
One consequence of galaxy mergers is the formation of a gravitationally bound binary SMBH system, which is expected to have a significant impact on the nuclear galactic dynamics and mass profile of the host \citep[e.g.,][]{merritt06,volonteri03}.  Post coalescence, the SMBHs from each galaxy are dragged towards the remnant's center by dynamical friction \citep[][]{chandrasekhar43}.  At separations of $\sim$10~pc, dynamical friction becomes less efficient at removing angular momentum, and other processes must drive binary orbital evolution.  Once the binary separation shrinks to $\lesssim 10^{-2}\;$pc, gravitational radiation becomes the dominant angular momentum loss mechanism and quickly drives the binary to coalescence \citep[e.g.,][]{begelman80}.  After the black holes coalesce, the remnant black hole may experience a kick due to the anisotropic emission of gravitational waves up to several 1,000~$\mathrm{km\ s^{-1}}$, and heavily depending on the spins and mass ratio of the precursor binary \cite[e.g.,][]{campanelli07}. 

The duration of the stage of binary evolution from $\sim$$10$ to $0.1$~pc is highly uncertain \citep{begelman80}, where the timescales range upwards of a few Gyr \citep{khan13} to longer than a Hubble time \citep{yu2002}, depending on the nuclear environment and properties of the binary.  This bottleneck of stalled binary SMBHs at $\sim$pc-scale separations is generally referred to as the "final pc problem".   However, the "final pc problem" may not be a problem after all \citep{milosavljevic03}, where many authors have recently suggested a number of different physical models involving the scattering of stars \citep[e.g.,][]{khan15} and interactions with a gaseous disk \citep[e.g.,][and references therein]{haiman09} to shrink binary separations within $\sim$10~Myr.  The observations of binary SMBHs at these separations ($\sim$~0.1$-$10~pc) is critical to assessing which of these physical models, and under what circumstances, is applicable.

The detection of gravitational waves emitted by binary SMBHs with orbital periods of a few weeks to a few decades (with corresponding gravitational-wave frequencies of nHz-$\mu$Hz) is the primary goal of Pulsar Timing Array (PTA) experiments \citep[][]{mclaughlin+13,hobbs+13,verbiest+16,nanograv12.5}. The detection of gravitational waves (in the range of 0.1--100~mHz) of less massive, coalescing SMBHs and intermediate mass black holes (IMBHs, M$_{\mathrm{BH}}\sim10^{4}-10^{7}$M$_{\odot}$) is sought in the 2030's by the planned Laser Interferometer Space Antennae \citep[LISA,][]{lisa17} mission. However, the multi-messenger science of PTAs and LISA strongly depends on our ability to identify electromagnetic observational constraints of the binary SMBH population in the local Universe \citep[][]{gwastro-review,witt+20,charisi+21}, and this is one of the prime motivations for this work. 
\subsection{The Search For Binary SMBHs}

There are two main techniques utilized in searching for binary SMBHs with expected separations $\mathrm{\lesssim10~pc}$ in electromagnetic observations.  The first is to look for quasi-periodic light curves, which are predicted to be a consequence of the binary's orbital motion  \citep[e.g.,][]{graham+15,charisi+16,liu.t.+16,liu.t.+19,liu.t.+20}.  There are a few physical models that give rise to this expected periodic signature in binary SMBH systems, including the periodic modulation of the accretion rate onto one (or both) of the SMBHs \citep[as is modeled for the famous blazar OJ~287,][]{pursimo+00}, the Doppler-boosting of a relativistically orbiting SMBH \citep[e.g.,][]{d'orazio+15}, and even gravitational self-lensing of the binary \citep[e.g.,][]{spikey+20}.  However, there are also a few competing physical models that can explain the light curve periodicities \textit{without the need for a binary SMBH}.  These include a precessing relativistic jet \citep[where the precession is not caused by the binary's orbital motion,][]{rieger04}, a warped accretion disk \citep{hopkins+10}, or simply stochastic variability misconstrued as truly periodic \citep[][]{vaughan16}.  The second major (electromagnetic) observational approach to identifying binary SMBHs is to search for the Doppler-shifting of quasar broad emission lines due to the binary's orbital motion \citep[e.g.,][]{bogdanovic+09,tsalmantza+11,Eracleous_12,ju+13,shenyue+13}, in exact analogy to spectroscopic searches for binary stars based upon Doppler-shifted stellar absorption lines.  


In this paper we study the 88 candidates identified by \cite{Eracleous_12} among 
15,900 quasars (all z $<$ 0.7) from the Sloan Digital Sky Survey (SDSS). Their defining characteristic is that their broad H$\beta$ emission-lines are offset from the host galaxy rest frame (as defined by the narrow lines) by velocities in excess of $\sim$1,000 $\mathrm{km\ s^{-1}}$. The rationale behind this selection is that the gas that makes up the broad-line region (BLR) is gravitationally bound to one of the black holes, whose orbital motion about the center of mass is responsible for the Doppler-shifting of the lines.

One clear prediction of this model is the oscillation of the broad line velocities about the host galaxy rest frame as the black holes complete their orbits.  This prediction was directly tested in \cite{runnoe17} via the use of multi-epoch spectroscopy.  In that study, several spectra were obtained per target, over intervals ranging from a few years to a decade to search for systematic velocity changes consistent with a binary SMBH scenario.  While several objects displayed promising radial velocity variations, the lack of radial velocity changes in the other targets does not rule out the binary SMBH hypothesis, since the expected orbital periods are on the order of a few decades to hundreds of years 
and it simply might take longer time baselines to detect their orbital motion via spectroscopic campaigns.
  
Alternatively, it is possible a \textit{recoiling} SMBH may be responsible for the broad-line shifts.  Recoiling SMBHs are believed to be the result of the coalescence of binary SMBHs, where (depending on the spin state and mass ratio of the precursor binary) the anisotropic emission of gravitational waves, as the binary coalesces, can lead to a velocity kick reaching values up to a few 1,000~$\mathrm{km\ s^{-1}}$ \citep[though a few $\times 100\; \mathrm{km\ s^{-1}}$ is a more likely outcome,][]{dotti+10,lousto+12,blecha+16}.  In this scenario, the BLR is again gravitationally bound to the recoiling SMBH, leading to the systemic Doppler shift of the BLR's emission lines.  


Finally, it is possible that an intrinsic (and one-sided) BLR outflow is responsible for the systematic Doppler shift of the offset emission lines seen in \cite{Eracleous_12}.  Such an outflow may result from the transfer of momentum by a relativistic jet, where jets in active galactic nuclei (AGN) are commonly shown to impart significant energy and momentum to ionized gas clouds in simulations \citep[][]{wagner+12} and in  observations of narrow line region gas \citep[][]{harrison+18,santoro+20}.  The motivation for the work described in this paper is to help explore these physical models by analyzing the observed mas-scale radio properties of these systems in conjunction with the optical data, and comparing with our expectations based on the competing physical models.  

In this paper we present the results of the VLBA X-band observations of 34/88 quasars from the \cite{Eracleous_12} sample of binary SMBH candidates.  These VLBA targets were chosen as they were the 34 radio-brightest quasars in the \cite{Eracleous_12} sample, as identified from a Very Large Array (VLA) "finder survey" aimed at identifying suitable radio candidates for further VLBA follow-up.  The results from those VLA observations are to be presented in the upcoming companion paper, \cite{sarah+21}.  In section~\ref{sec:data_analysis}, we present the data reduction methodology and procedures utilized.  In section~\ref{sec:results}, we present the major results of these analyses, with their interpretations and implications discussed in section~\ref{sec:discussion}.  Finally, in section~\ref{sec:conclusion} we summarize the main findings of our work.  Throughout this paper we adopt a $\Lambda$CDM cosmology, with H$_{0}=67.74$ km s$^{-1}$ Mpc $^{-1}$, $\Omega_{\lambda}=0.69$, and $\Omega_{m}=0.31$ \citep{planck+16}.  

\section{VLBA Observations and Data Analysis} \label{sec:data_analysis}
The 34 targets that make up our sample are listed in Table~\ref{table:image_properties}. Following \cite{Eracleous_12}, we refer to the targets for the rest of the paper using the first six digits of the right ascension, i.e. to integer seconds (e.g., Jhhmmss). This convention identifies them uniquely in Table~1 of \cite{Eracleous_12}, where more precise coordinates of the SDSS sources can be found. The targets were observed with the VLBA at X-band (centered at 8.5~GHz), between the dates of February 13, 2014 and July 14, 2014 (Program ID: BS231).
The targets were observed with varying hour angles in order to maximize $u$-$v$ plane coverage and with integration times designed to achieve root mean square (RMS) image intensity noise levels $\sim20$ times below the X-band flux densities detected previously at the VLA \citep[][]{sarah+21}, extrapolated from the 10~GHz VLA observing frequency and assuming flat spectral indices.
This requirement was intended to allow for 10$\sigma$ detections of point sources, assuming the unresolved VLA radio sources ($\sim0.1\arcsec$ restoring beams) are split into two equal-flux, compact components on VLBA mas scales.  

The experimental setup included full polarization capabilities, 2~s integration times, and a total bandwidth of 256~MHz, split into eight spectral windows (with 256 channels per spectral window).  Phase-referencing observations were employed, as absolute positional information was necessary for the purposes of further astrometric comparisons with other wavelengths \citep{beasley95}.  The phase calibrators were all within a few degrees of the science targets, and the cycle time used  for the "nodding" mode observations (phase-target-phase scans) was $\sim$5.5~min (2 min on phase calibrators, 3.5 min on science targets). We used a single AGN from the following list as both our fringe finder and bandpass calibrator for each observing session: 3C~345, 3C~84,4C~39.25, and 3C~273.

We calibrated the data in AIPS \citep{moorsel+96} using the standard calibration procedures applied in the pipeline task \texttt{VLBARUN} for continuum imaging, where appropriate reference antennae were chosen (typically Pie Town, unless an issue was present with that station).  Log-based flagging was performed prior to calibration.
After calibration, we used CASA \citep{mcmullin+07} for further flagging of radio-frequency interference (RFI), combination of science target visibility data from multiple different observing sessions with the task \texttt{concat}, and then cleaning and imaging with the task \texttt{tclean}. We used a natural weighting scheme for source finding and imaging (as this has the greatest sensitivity for point sources), with cell sizes of $\sim$4 pixels per restoring beam.  We also made wide-field images to search for any cores on scales up to $\sim$0.9\,arcsec.  This was accomplished with the \texttt{wproject} gridding algorithm \citep[][which corrects for the effect of non-coplanar baselines]{cornwell+08}, with the use of \texttt{wprojplanes=-1} in CASA.  This choice automatically determines the number of planes to use based upon your data and image size\footnote{CASA determines this number based upon the following formula: $\mathrm{N_{wprojplanes}=0.5\times\frac{W_{max}}{\lambda}\times\frac{imsize}{(radians)}}$.  Here,  imsize is the image size, $\lambda$ is the wavelength, and $\mathrm{W_{max}}$ is the maximum w in the uvw data (i.e., physical extent of the visibility data that is orthogonal to the image plane).}.

We determined the VLBA-based radio positions (given in Table~\ref{table:image_properties}) by fitting the image source regions with elliptical Gaussian functions, utilizing the CASA task \texttt{imfit}.  The resulting positional errors from fitting with \texttt{imfit} were added in quadrature with the phase calibrator position errors \citep{gordon+16}, and the errors associated with the phase-referencing technique itself \citep[e.g.,][on the order of 0.1~mas for all of our sources]{pradel+06}.  

\def\arraystretch{1}%
\setlength\tabcolsep{3pt}
\begin{deluxetable*}{lccccccccccr}[th]	\tablecaption{\label{table:image_properties} VLBA Source Detection/Image Properties \& Binary Model Constraints}
	\tablecolumns{12}
	\tablewidth{0pt}
		\tablehead{
		&&Right&&Image&Peak&Flux&a$_{\mathrm{proj}}$\tablenotemark{\small c}&&&&\\
		Source& z&Ascension&Declination&Noise Level&Brightness&Density\tablenotemark{\small b}&Limit&$\mathrm{B_{maj}}$\tablenotemark{\small d}&$\mathrm{B_{min}}$\tablenotemark{\small d}&P.A\tablenotemark{\small d}&P\\
		Name\tablenotemark{\small a}&&(J2000)&(J2000)&(mJy beam$^{-1}$)&(mJy beam$^{-1}$)&(mJy)&(pc)&(mas)&(mas)&($^{\circ}$)&}
	\startdata
	J022014&0.214&...&...&0.079&...&$<$0.394&...&2.82&1.73&$-$5.98&1\\
	J075403&0.273&...&...&0.064&...&$<$0.318&...&2.23&1.13&$-$11.2&1\\
	J082930&0.321&08 29 30.60017(2)&+27 28 22.6788(2)&0.013&0.254&0.306&$<$11.3&2.35&1.06&$-$6.67&1\\
	J091833&0.452&...&...&0.022&...&$<$0.110&...&3.02&1.76&$-$20.6&1\\
	J093100&0.460&...&...&0.030&...&$<$0.152&...&2.92&1.32&10.9&0.83\\
	J093653&0.228&09 36 53.84274(1)&+53 31 26.7936(2)&0.026&0.367&0.395&$<$8.16&2.28&0.907&$-$2.2&1\\
	J093844&0.171&...&...&0.020&...&$<$0.098&...&2.65&1.13&$-$2.11&1\\
	J094603&0.220&...&...&0.113&...&$<$0.565&...&2.94&0.951&$-$8.36&1\\
	J095036\tablenotemark{$\dagger$}&0.214&09 50 36.75657(2)&+51 28 38.0651(4)&0.575&6.07&24.6&$<$11.6&3.23&2.20&$-$11.1&0.31\\
	J095539\tablenotemark{$\dagger$}&0.259&09 55 39.82672(1)&+45 32 17.0044(3)&0.218&2.8&6.2&$<$10.9&2.64&1.12&$-$11.1&0.48\\
	J102106\tablenotemark{$\dagger$}&0.364&10 21 06.04639(2)&+45 23 31.8208(3)&0.978&8.05&60.9&$<$27.3&5.23&3.11&$-$9.42&0.07\\
	J105203\tablenotemark{$\dagger$}&0.400&10 52 03.17204(1)&+24 05 04.9758(3)&0.618&11.0&26.0&$<$16.5&2.98&1.13&$-$6.76&0.09\\
	J113330\tablenotemark{$\dagger$}&0.511&11 33 30.29867(1)&+10 52 23.3530(3)&0.018&0.171&0.567&$<$16.4&2.58&2.11&$-$6.48&1\\
	J115158&0.170&...&...&0.028&...&$<$0.138&...&2.61&1.12&$-$0.587&1\\
	J115449&0.470&...&...&0.133&...&$<$0.665&...&2.83&0.995&$-$8.03&0.23\\
	J121113&0.295&12 11 13.98237(2)&+46 47 11.9767(3)&0.063&1.09&1.64&$<$11.3&2.48&0.993&$-$5.94&1\\
	J122811\tablenotemark{$\dagger$}&0.321&12 28 11.88763(3)&+51 46 22.7816(5)&0.185&1.79&3.71&$<$12.2&2.54&0.96&$-$3.60&0.37\\
	J123001&0.448&12 30 01.028639(9)&+33 59 01.3926(2)&0.021&0.175&0.268&$<$14.2&2.33&1.08&$-$4.91&1\\
	J125142&0.189&...&...&0.025&...&$<$0.124&...&3.11&2.51&6.72&1\\
	J125809&0.310&12 58 09.313783(7)&+35 19 43.0153(2)&0.024&0.240&0.349&$<$9.96&2.11&0.780&$-$1.79&1\\
	J134617&0.117&...&...&0.134&...&$<$0.670&...&2.78&0.918&$-$9.33&1\\
	J140700&0.077&...&...&14.9&...&$<$74.5&...&3.07&0.922&$-$5.45&0.13\\
	J143455&0.176&...&...&0.135&...&$<$0.675&...&3.19&0.956&$-$6.74&1\\
	J151132&0.281&15 11 32.537730(8)&+10 09 53.0205(2)&0.214&6.23&7.87&$<$13.1&2.93&1.85&$-$15.95&0.42\\
	J151443&0.371&15 14 43.068283(8)&+36 50 50.3565(2)&1.22&21.0&54.1&$<$12.0&2.26&0.973&$-$4.90&0.06\\
	J152942&0.150&...&...&0.294&...&$<$1.47&...&2.44&0.978&1.68&1\\
	J153636&0.389&15 36 36.22321(1)&+04 41 27.0691(2)&0.026&0.568&0.789&$<$16.7&3.06&1.15&17.9&1\\
	J154340&0.421&...&...&0.447&...&$<$2.24&...&3.01&1.10&15.0&0.10\\
	J155654&0.165&15 56 54.47450(1)&+25 32 33.5967(3)&0.214&2.70&5.60&$<$7.31&2.51&1.03&$-$2.09&1\\
	J160536&0.441&...&...&0.504&...&$<$2.52&...&2.29&0.985&$-$7.38&0.08\\
	J162914&0.312&16 29 14.09282(2)&+15 14 15.3628(4)&0.015&0.086&0.121&$<$14.1&2.98&1.40&17.9&1\\
	J163020&0.395&...&...&0.135&...&$<$0.675&...&2.59&1.09&$-$5.88&0.32\\
	J171448&0.180&...&...&0.039&...&$<$0.195&...&3.01&1.25&7.72&1\\
	J180545&0.386&18 05 45.11318(1)&+22 51 37.8038(2)&0.215&4.53&5.38&$<$12.0&2.22&0.859&1.41&0.23\\ 
	\enddata
	~ \\
		\tablenotemark{$\dagger$}{These sources have partially resolved structure as determined by the CASA Gaussian fits
		.}\\
	    \tablenotemark{\small a~}{We refer to the targets by the first six digits of right ascension, following the convention of \cite{Eracleous_12}.} \\
		\tablenotemark{\small b~}{5$\sigma$ flux  density upper limits are given for non-detected sources.}\\
		\tablenotemark{\small c~}{We present upper limits on the instantaneous projected binary separation (under the assumption that there is actually an unresolved binary within our source detections), using the major axis of the restoring beam as our limiting angular separation (or the major axis of the deconvolved image component fit for the partially resolved sources).  See section ~\ref{binary_discussion} for a more detailed discussion of this limit. } \\
		\tablenotemark{\small d~}{$\mathrm{B_{maj}}$ and $\mathrm{B_{min}}$ refer to the major and minor axes of the restoring beam, respectively (except for partially resolved sources where we report the deconvolved image component axes instead), and P.A refers to the beam position angle.}
\end{deluxetable*}
\section{Results} \label{sec:results}
\subsection{Source Detections and Properties}
\begin{figure*}[]
	\vspace{20pt}
	\begin{center}
		\includegraphics[width=6.5in]{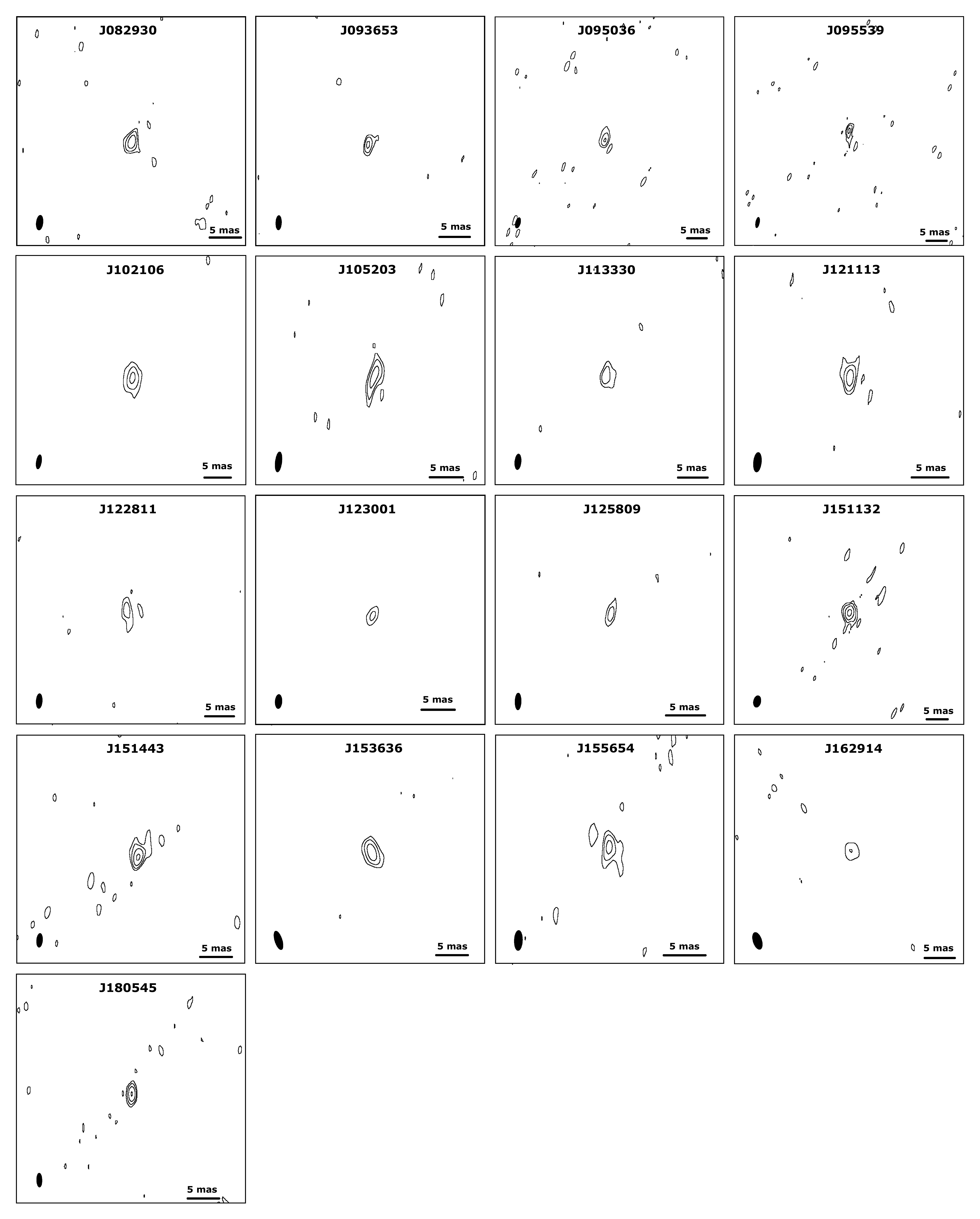}
	\end{center}
	\caption{\label{fig:contour_maps} Intensity contours for the 17 targets detected in our VLBA campaign.  Contours start at~3 times the RMS noise level in the image (Table~\ref{table:image_properties}, and are spaced by factors of 2 thereafter.  The restoring beam shapes are shown in the bottom left corners of each image as a filled ellipse, and 5~mas angular scale bars are shown in the bottom right.}  
\end{figure*}

We detect 17 sources in our VLBA imaging, corresponding to exactly half of our observed target sample (17/34).  Our source detections consist primarily of unresolved point sources, with six sources determined to be partially resolved by our elliptical Gaussian fits in CASA\footnote{A few sources were found to be partially resolved by these fits with deconvolved image component sizes smaller than the synthesized beam sizes; for these we reject the presumption that our images resolved any stucture on the grounds that this would be unphysical (and assume this is just an error associated with the the Gaussian fit algorithm itself).} (these sources are designated in Table~\ref{table:image_properties}).
We present the image contours for the 17 VLBA sources for which we claim detections in Figure~\ref{fig:contour_maps}, where we define a formal detection limit of 5$\sigma$ (base contours of 3$\sigma$, spaced by factors of 2). The image RMS noise level, detection flux density and intensity, half power beamwidth (HPBW) angular size and position angle of the restoring beam axes (except for partially resolved sources where we report the HPBW of the deconvolved image component sizes instead), and sky positions are given in Table~\ref{table:image_properties} (note that for undetected sources we also present 5$\sigma$ flux density upper limits).  

In exploring our non-detections, we found that five targets had issues with poor data quality (i.e., low number of observing antennae and persistent RFI) that led to sensitivities that were more than a factor of two worse than than our target sensitivity (as defined in \S\ref{sec:data_analysis}).  Given that we know all of our targets are radio-emitting at a level of at least 80~$\mu$Jy when observed on arcsecond scales, it is likely that many of our 17 undetected sources are simply dominated by radio emission on a larger scale (e.g.~broad lobes from a very young radio jet or compact symmetric object, or perhaps the radio emission in those objects observed at low resolution is dominated by star-formation). 
\begin{figure}[]
	\centering
	\includegraphics[scale=0.45]{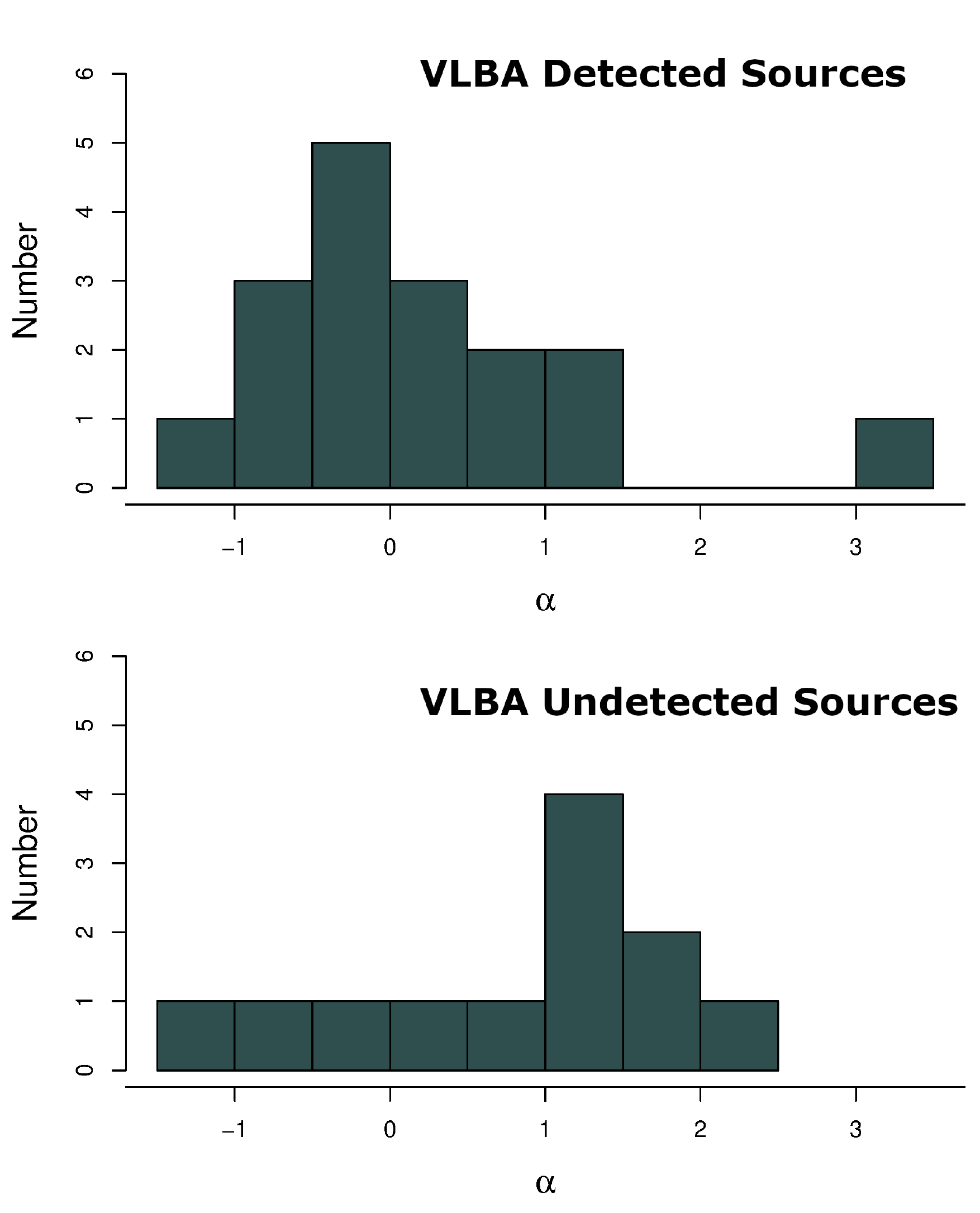}
	\caption{Spectral index, $\alpha$, histograms for the detected source population (top) and the undetected source population (bottom).  The spectral index was determined from the X-band VLA observations to be presented in \cite{sarah+21} (note that we do not have large enough bandwidths with the VLBA observations to reliably measure spectral index information). We have excluded from these plots the five VLBA-undetected sources for which the measured RMS was more than a factor of two worse than our design sensitivity due to RFI or array issues.
	}
	\label{fig:hist}
	\vspace{-3mm}
\end{figure}
However, they are resolved by our VLBA observations and do not exhibit any detectable compact core, where any Fourier components corresponding to large-scale structures intermediate between milliarcsecond and arcsecond scales may be filtered out of the VLBA visibilities due to the VLBA's lack of small-distance baselines.  We find some support for this possibility when looking at the distribution of the arcsecond-scale radio spectral indices of detections and non-detections, shown in Figure~\ref{fig:hist}; here, the spectral index, $\alpha$, is defined as $F_{\nu}\propto\nu^{-\alpha}$, where $F_{\nu}$ is the flux density, and $\nu$ is the frequency.\footnote{A full presentation of the VLA finding survey and spectral index information for the entire \cite{Eracleous_12} sample will be presented in \cite{sarah+21}}
The errors on the indices shown in Figure~\ref{fig:hist} are all below $\sim0.3$, with most being on the order of 10\%. As may be expected, the sources with detected compact VLBA components have a significant cluster around $\alpha\sim0$, consistent with objects dominated by synchrotron self-absorbed cores. Meanwhile, our non-detections are dominated by targets with relatively steep ($\alpha\gtrsim0.5$) spectra, consistent with diffuse synchrotron emission from lobe-type structures \citep[e.g.,][]{tremblay+16}, or star-forming galaxies \citep[e.g.,][]{klein+18}.  
For the four remaining flat/rising spectrum sources, we hypothesize that flux density variability may play a role in our non-detections with the VLBA (where the time between VLA and VLBA observations was roughly a year).

\subsection{Astrometric Offsets}\label{sec:astrometry}
A major motivation for these VLBA observations was the benefit from the VLBA's precise mas-scale astrometry.  We compared our fitted positions for the VLBA-detected targets with the optical source positions inferred from both the Sloan Digital Sky Survey (SDSS) and \textit{Gaia}~DR2 catalogs, all tied to the International Celestial Reference System (ICRS).  The positional uncertainties for SDSS include both the centroid and astrometric calibration errors (typically on the order of a few mas and $\sim 45$--75~mas respectively, \citealt{pier+03}).  The \textit{Gaia} celestial reference frame (GAIA-CFR2) is nominally aligned to the ICRF3 (the current instantiation of the ICRS, as defined by the VLBA, \citealt{gordon+16,charlot+20}) to $\sim$20-30 $\mu$as \citep{mignard+18}.  The \textit{Gaia}~DR2 positions are the full g~band photometric centroids and the position errors are determined from the five-parameter astrometric solution, as described further in \cite{lindegren+18}.  

\begin{figure}[t]
	\centering
	\includegraphics[scale=0.45]{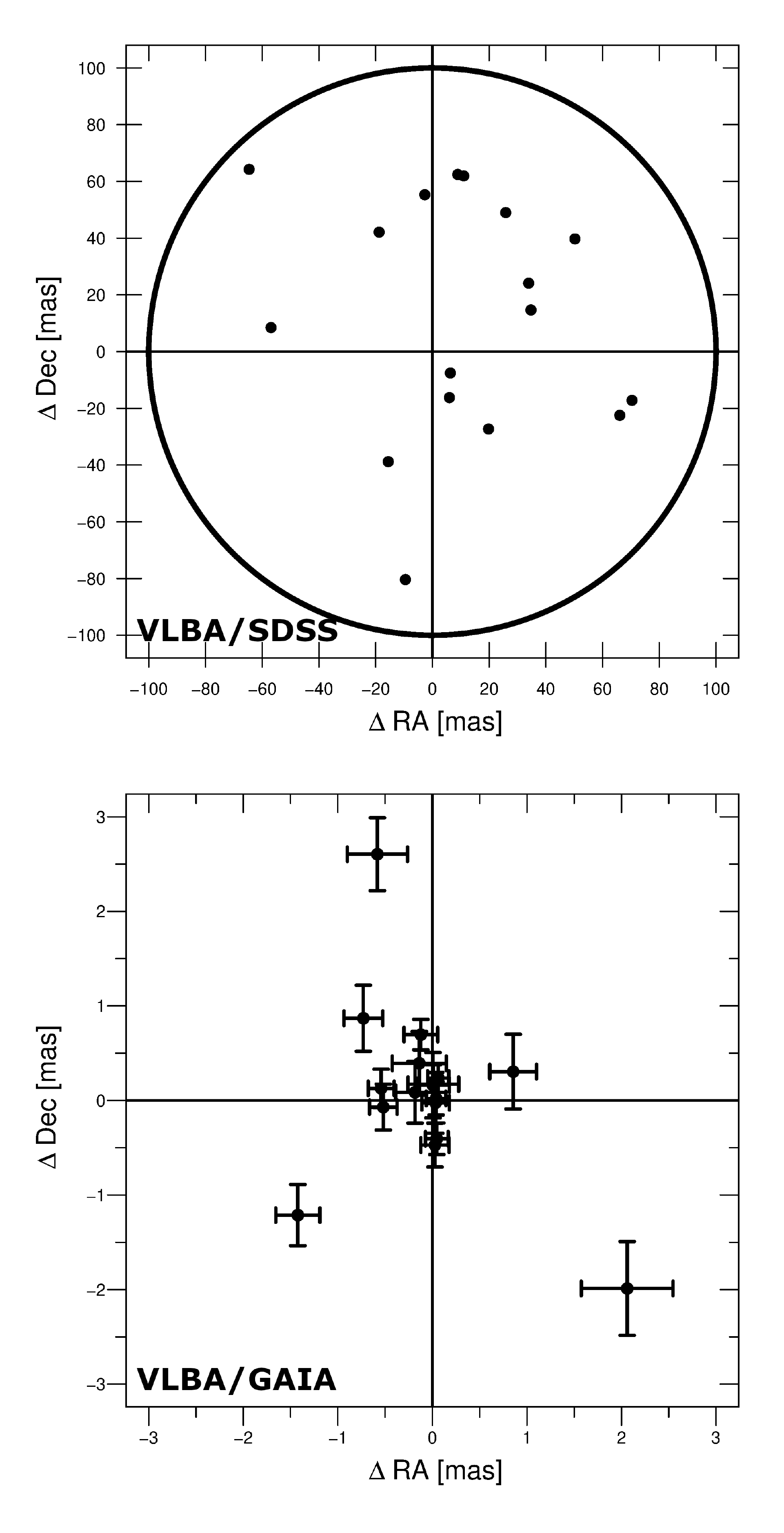}
	\caption{ \textit{Top}:  Positional offsets between the SDSS optical and VLBA radio source positions are plotted as black points.  The black circle shows the lowest 95\% random offset error associated with any one source (the greatest error associated with any one source is $\sim40$~mas greater than this smallest error radius).  \textit{Bottom}:  Positional offsets between the \textit{Gaia}~DR2 optical and VLBA radio source positions are plotted as black points, with associated 1$\sigma$  errors as described in Section \ref{sec:astrometry}.
	}
	\label{fig:offsets}
	\vspace{-3mm}
\end{figure}

In Figure~\ref{fig:offsets}, we show the resulting VLBA-SDSS and VLBA-\textit{Gaia} radio/optical sky position offsets (note that the errors in these offsets are determined by adding all of the associated errors for the VLBA/\textit{Gaia} and VLBA/SDSS positions in quadrature, respectively).  For the VLBA-SDSS offsets, we show in the black circle the smallest 95\% random error circle associated with any of our sources (due to the errors being greater than the offsets for all of the sources), where the radius of this ``smallest error circle'' represents the least stringent (i.e., smallest error) associated with any of our VLBA-SDSS offsets.  Clearly, none of the VLBA-SDSS offsets are discrepant at the $>$95\% confidence level, and the SDSS astrometry is not sufficient to find any statistically significant offsets.  
We also show the VLBA-\textit{Gaia} offsets with 1$\sigma$ error bars, where we find three sources with $>3\sigma$ offsets: J095036, J121113, J122811.  The corresponding projected physical offsets (i.e., in the plane of the sky) are 9.6~pc, 8.6~pc, and 13.9~pc, respectively.

\section{Discussion} \label{sec:discussion}

\subsection{The Binary SMBH Scenario} \label{binary_discussion}

The clearest evidence for a binary SMBH in high-resolution radio images would be to find two compact (unresolved but clearly separated) objects. Given the single point sources in all our detections, we thus find no complete evidence for or against binary SMBHs from the present data. 
Assuming our sources actually contain binary SMBHs, there are two ways in which our VLBA observations failed to yield double radio source detections in our imaging: namely, the second black hole is too faint to be detected or the binary remains unresolved by our observations.  We further discuss these two possibilities below.

First, there could be a second, radio-quiet companion SMBH in the system undetected by the VLBA at X-band due to being under-luminous at this frequency, but still existing outside of the VLBA angular resolution limits. Such a radio-quiet companion could be present, and may be experiencing lower (or non-existent) levels of accretion (or perhaps a lower black hole spin), leading to a lack of relativistic jet formation that prevents the secondary from being a luminous radio emitter \citep[e.g.,][]{tchekhovskoy+11}. Another scenario which might give rise to an under-luminous companion is emission that is relativistically beamed away from us due to a large angle of the jet to the line-of-sight.  Given that only approximately 1\% of field galaxies support strong radio emission, this is a plausible scenario.  

Following the methodology presented in \cite{sarah+11}, we tested the possibility of an undetected radio-faint companion SMBH by using a flat-spectrum, redshift-dependant radio luminosity function of AGN to assess the probability that said companion would remain undetected in the flux-density limits of our VLBA observations of detected sources.  Our analysis assumes the ``pure luminosity evolution model'' for the luminosity distribution of radio-loud galaxies given by \cite{dunlop+90}, and accounts for the full range of Doppler-boosting and intrinsic luminosities of observed AGN.  The use of this luminosity function assumes that the merging of galaxies, and binary SMBH nature for this companion has no impact on its radio luminosity.  This assumption is reasonable, as the fraction of black holes manifesting themselves as AGN has been shown to rise systematically with the proximity of galaxy pairs, implying that ongoing interactions can increasingly trigger AGN \citep{ellison+11}, and even specifically radio-loud AGN \citep[][]{chiaberge+15}. Thus, if these are indeed systems that contain a binary SMBH, they likely have a higher chance of both being radio-emitting AGN, and it is not clear what processes would give rise to the quenching of a companion's radio emission in a binary system.  We present the resulting probabilities, P, that we would have detected a radio-luminous companion if it existed outside of the VLBA angular resolution limits in Table~\ref{table:image_properties}.  For this calculation, we assumed flux density limits of 5$\sigma$, and used the SDSS-measured redshifts (also given in Table~\ref{table:image_properties}).  We also give the probabilties we would have detected a radio luminous SMBH for the undetected VLBA sources as well (i.e., sources where we report no VLBA detctions; though see section ~\ref{sec:results} for a thorough discussion of our non-detections). As is evident from Table~\ref{table:image_properties}, this argument suggests we should have been sensitive enough to detect some of the radio-emitting companion SMBHs, if they exist and are outside of our VLBA angular resolution limits.  However, there are clearly several sources where deeper observations are necessary to draw any firm conclusions.  

Secondly, it is possible that there is a binary SMBH or even a binary AGN, but it is simply at a projected orbital separation within the limits of our observing resolution.
Utilizing our X-band VLBA images, we present limits (given in Table~\ref{table:image_properties}) on the instantaneous, projected orbital separation of the binary based on the assumption that a binary's projected separation is smaller than the observing resolution we reached.
These limits are presented under the assumption that both black holes are radio luminous, and would have been bright enough to be detected in our radio imaging if they were resolvable by our observations.  The instantaneous, projected separation is the projected separation of the binary at a given orbital phase, and therefore represents a lower limit on the true orbital separation of the binary.  If the orbital phase and inclination of the binary could be determined, this would allow for the  translation of this quantity into an actual measurement of binary orbital separation \citep[][]{nguyen+20}.  If strong constraints could be placed on orbital phase and inclination, this would allow for meaningful upper limits  to be obtained on the binary total mass and period \citep[see equations 1 and 2 from ][]{Eracleous_12}.

The expected binary separations of the \cite{Eracleous_12} sample, as modeled by \cite{pflueger18}, are on the order of 0.1-10~pc.  
Therefore, higher angular resolution observations would provide a more stringent test of the binary SMBH hypothesis.  Future follow-up with the VLBA at higher frequencies will help us to probe the binary hypothesis down to smaller angular and physical scales (i.e., Q-band, which is a frequency of $\sim$~43~GHz, and above will increase the angular resolution by greater than $\sim$~5 times the X-band value).  However, higher-frequency observations with the VLBA are less sensitive, and require more restrictive weather conditions in order to employ the requisite observational setup for proper calibration.  Thus, it may be desireable to observe the brightest quasars from our sample with Q-band (or even higher frequency) VLBA observations.  We can also achieve higher angular resolution by increasing the maximum baseline of our data set by the inclusion of data from the European VLBI Network (EVN) or the Atacama Large Millimeter/sub-Millimeter Array (ALMA, where we would need to use a higher-frequency VLBA data set in combination with ALMA, since ALMA's lowest frequency band extends from 35-50~GHz).  Expanding beyond the VLBA, the Event Horizon Telescope (EHT) would help us to probe the binary SMBH scenario down to 25$\mu$as scales \citep{eht_paper}, representing a $\sim100\times$ improvement in angular resolution from the VLBA observations presented here (effectively probing physical scales down to $\sim$~0.1~pc, although the EHT's relatively poor sensitivity combined with a likely falling flux density at high frequencies might mean very long integration times).  Similarly, the next generation Very Large Array (ngVLA) will be able to probe angular scales down to $\sim$3~mas$-$80~$\mu$as for the respective frequency range of 2.4$-$93~GHz \citep{Murphy}.  If we are able to eventually resolve a binary SMBH and subsequently track its orbital motion with very long baseline interferometry \citep[VLBI, as was done for the binary SMBH 0402+379 by][]{bansal+17}, we would have leverage on an entirely new and independent method of black hole mass determination and measurements of the Hubble constant \citep[to within $\sim$30\% and $\sim$10\% respectively,][]{d'orazio+18}.  

Finally, deeper (i.e., higher sensitivity) high-dynamic-range imaging with the VLBA may be able to reveal the existence of a pc-scale jet.  Radio jets associated with AGN tend to have a radio spectral index of $\alpha\sim0.7$, so lower-frequency imaging may help in detecting their emission (i.e., C-band, which is a frequency of $\sim$~6~GHz, and is roughly twice as sensitive as X-band for the VLBA).  For reference, a typical core-to-jet-knot flux density ratio is on the order of a few for radio-loud AGN \citep[e.g., see Figure~7 from][]{jorstad+17}.  Therefore we would expect a $\sim10\times$ increase in integration time to detect any jet components at X-band frequencies.  However, given that C-band is roughly twice as sensitive as X-band for the VLBA, and the flux density likely increases at lower frequencies for these jet components, we expect a factor of $\sim$ a few increase in integration time necessary to detect such components with C-band VLBA observations.  It is also possible jet precession due to a binary SMBH may result in a curved pc-scale jet morphology, as was seen in the AGN KISSR~434 \citep{kharb+19}.  Multi-epoch imaging with the VLBA may be able to monitor such a jet precession due to the periodic increase in luminosity as the jet swings into our line-of-sight and relativistic beaming effects become significant \citep[as is modeled in the famous AGN OJ~287, BL~Lacertae, and 3C~66B,][]{valtonen13,stirling03,sudou03}. 
\subsection{The Recoiling SMBH Scenario}

After a binary SMBH system exhausts all of its orbital energy in the form of gravitaional waves, it coalesces and the remnant black hole may experience a recoil due to the anisotropic emission of gravitational waves \citep{peres62,bekenstein73}.  In the recoiling SMBH scenario, the large ($\sim$~a few~1000~$\mathrm{km\ s^{-1}}$) velocity offsets of the broad emission lines seen in \cite{Eracleous_12} are due to the Doppler shifting of emission lines by a BLR that is gravitationally bound to the recoiling SMBH.  Thus, the velocity offset of the broad lines would represent a lower limit on the recoiling SMBH velocity (since we are only observing the radial velocity component from the offset broad emission lines, and dampening in the galactic environment works to slow the SMBH's movements over time).  If this scenario is correct, it also implies very large recoil velocities \citep[where, for a large portion of the initial conditions parameter space, these kicks are typically on the order of a few hundred $\mathrm{km\ s^{-1}}$ in simulations, ][]{dotti+10,lousto+12,blecha+16}.
The maximum recoil velocity identified in fully relativistic simulations of black hole mergers is up to $\sim$5,000~$\mathrm{km\ s^{-1}}$ \citep[e.g.,][]{campanelli07,lousto+19}.  In general, the recoil velocity is larger for equal mass-ratios \citep[][]{herrmann+07}, and near-maximal black hole spin magnitudes, with spin vectors either both partially aligned with the orbital angular momentum vector (in the so-called "hangup kicks") \citep[e.g.,][]{lousto+11}, or anti-aligned and lying in the orbital plane (in the so-called "super kicks") \citep[e.g.,][]{gonzalez+07}.  Taking into consideration the above constraints on recoil velocities, the quasars with Balmer lines exhibiting the greatest velocity offsets from \cite{Eracleous_12} (i.e., those showing offsets approaching or greater than 5000 $\mathrm{km\ s^{-1}}$) are unlikely to be recoils.  However, this certainly does not rule out the  recoil scenario for the black holes with more moderate velocity offsets. 

The observation of significant positional offsets (on the scale of many tens of pc to kpc scales) between the radio core of the SMBH and the host galaxy's dynamical center would constitute strong evidence for the recoil scenario.  In Figure~\ref{fig:offsets}, we show the positional offsets between the VLBA and the optical source positions from the \textit{Gaia}~DR2 and SDSS optical survey catalogs.  The optical positions measured from these surveys trace the position of the quasar's accreting SMBH, making it difficult to ascertain the true optical photometric centers of the host galaxies.  High angular resolution observations with an optical observatory of sufficient sensitivity, high dynamic-range, and a sufficiently stable and well-characterized point spread function function (PSF) would allow for the quasar core subtraction from the imaging, and precise modeling of the host's photometric center.  The radio/optical offsets ascertained from the comparison of our VLBA positions with the type of optical imaging and modeling described above could be considerably larger,  and potentially much more illuminating.  

However, we do measure statistically significant ($>3~\sigma$) offsets on the scale of $\sim$10~pc for three of our targets.  We can explain these offsets in a few different ways:

\begin{enumerate}
	\item If the \textit{Gaia} optical position is dominated by thermal radiation from the quasar accretion disk, we may be seeing the offset beween the radio core and accretion disk from the same AGN.  One common interpretation for the radio core is that it represents the location of a standing recollimation shock of the AGN's relativistic jet, several-to-tens of pc downstream of the central engine \citep{daly88}.  Therefore, this scenario is completely consistent with a single AGN, but does not rule out any of the proposed physical models for the broad emission line offsets.      
	\item It is possible that unresolved optical jets are responsible for the radio/optical offsets seen in our  sample, where similar-scale offsets were reported by \cite{petrov+17} for a sample of VLBA sources cross-correlated against the \textit{Gaia} DR2 catalog.  The idea in this scenario is that while the VLBA is sensitive to compact source structures on sub-mas scales (i.e., the radio core), \textit{Gaia's} centroiding position is more heavily weighted by extended low surface-brightness emission, which can skew the results. However, if the \textit{Gaia} source position is heavily skewed by an optical quasar jet, the quasar optical spectrum should be dominated by non-thermal emission from the jet (or at least show a significant non-thermal contribution), in contrast to a spectrum well fit by a thermal disk emission model (i.e., the ``big blue bump'').  In principle, this should be testable with spectral energy distribution (SED) fitting.
	\item It is possible that we are seeing the "core shift" of some quasar jet, due to its observation at different frequencies.  This scenario, like the previous one, implies a more radiatively inefficient accretion mode for these quasars (i.e., these would be ``jet mode'' systems), where the jet is dominating the quasar's light output as opposed to the accretion disk. 
	The idea in this scenario is that the radio core is actually the $\tau=1$ location at which the jet becomes optically thick to synchrotron self-absorption, which in principle will depend on the frequency \citep[e.g.,][]{hada+11}.  In this scenario, higher-frequency observations will reveal a core location which shifts \textit{towards} the black hole (if a resolved VLBI jet were to be detected, this could help delineate the direction of the black hole).  Similar to the previous scenario, this hypothesis implies a non-themal optical spectrum which should be testable via SED-fitting.   
	
	\item Another scenario which could account for the VLBA-\textit{Gaia} offsets is a recoiling SMBH displaced from its host's center on kpc scales, where the $\mathrm{\sim1000-1800\ km\ s^{-1}}$  range of broad-line velocity offsets for these three quasars would be extreme, but plausible  (line-of-sight) recoil velocities.  The optical centroid of the \textit{Gaia} source positions is \textit{mostly} traced by the light emanating from the quasar accretion disk.  However, there is likely a very small, but nonetheless salient, contribution from the host galaxy starlight which acts to displace the optical centroid from the quasar position.  This optical photocenter shift might be able to account for the measured radio-optical offsets, assuming the radio position specifies the SMBH location.  
	However, the magnitude of this optical photocenter shift will depend on the specific host galaxy light profile, instrumental PSF shape, and relative contrast between the quasar and host galaxy.  Thus, this kind of analysis is beyond the scope of the present study, and we defer a more in-depth treatment to a future work.  One  consideration which lends credence to the recoil scenario is that kpc-scale offset recoiling SMBHs with $\mathrm{\sim1000-1800\ km\ s^{-1}}$ velocities are expected to be long-lived (i.e., $\sim1-100$ Myr) and accreting at relatively low rates \citep[e.g.,][]{blecha+11,sijacki+11}, leading to low AGN optical luminosities and a higher contribution of host galaxy light to the optical photometric center.  
	If these systems truly contain recoiling SMBHs, in principle there should be proper motion of the recoiling SMBH away from the host  galaxy's center, and decelerating due to the SMBH overcoming the host galaxy's gravitational potential well (and moving through the interstellar/intergalactic medium).  One could hope to detect such proper motions with a radio observatory of sufficient angular resolution and astrometric precision.  In practice, this is likely outside the scope of possibilities for the VLBA, or any other planned radio observatory in the near future.  For reference, we estimate that the VLBA could detect a 5$\sigma$ astrometric shift for the nearest object in our sample in observations separated by $\sim$35 years, assuming a 4,000~$\mathrm{km\ s^{-1}}$ tangential velocity component (i.e., in the plane of the sky).  We must stress that such high recoil velocities are not expected and would represent highly improbable events.  It is therefore extremely unlikely we will be able to measure any astrometric shifts induced by recoiling SMBHs in these quasars with any current or upcoming facillities.  
	\item Finally, it is possible we are actually seeing two different AGN, one radio bright and the other radio-quiet, but optically bright (see \citealt{orosz+13} for a discussion on how radio-optical offsets could imply the existence of dual AGN).  This scenario is a natural outcome of the situation where major galaxy mergers in gas rich systems lead to SMBH fueling and AGN activity as the SMBHs are driven to coalescence (and higher gas densities near the galactic center, see e.g., \citealt{barrows+18}).  High angular resolution optical imaging, combined with high-resolution X-ray imaging may be able to lend credence to this scenario.  
\end{enumerate}

\subsection{ Interactions with a Relativistic Jet}

 The final scenario we consider for producing the observed broad emission line offsets is the transmission of momentum from a relativistic jet to the BLR gas clouds, either via gas entrainment or ram pressure by the jet.  One prediction from this scenario is that the jets more closely aligned to the line-of-sight should yield higher measured velocity offsets as they would have a higher proportion of radial-to-tangential velocity components.  Similarly, jets with smaller angles to the line-of-sight should have higher inferred luminosities due to relativistic beaming effects \citep{blandford79}.  However, we did not see any evidence for a correlation between VLBA 8.5~GHz luminosities and broad line velocity offsets, albeit for the small sample size of 17.  The bulk of the evidence against this scenario will be presented in \cite{sarah+21} for the full sample of radio-detected \cite{Eracleous_12} objects.  
 
 Another way to distinguish between this model and the recoiling or binary  SMBH scenarios is by observing the Fe~K${\alpha}$ lines (should they exist) of the \cite{Eracleous_12} sources with a suitable high energy-resolution X-ray spectrometer.  
The broad (full width at half maximum of $\sim$~a few to tens of thousands $\mathrm{km\ s^{-1}}$) Fe~K$\alpha$ line is a 6.4~keV K-shell fluoresecnce line, believed to originate from the reprocessing of the X-ray continuum by matter within the inner regions of the accretion disk when observed in AGN \citep[e.g.,][]{george91,fabian89}.  The line-broadening is a combination of Doppler and general relativistic effects.\ \ 
Since the broad Fe~K$\alpha$ line emission is tied to the accretion disk, its rest frame is also inexorably tied to the SMBH.  If the broad-line shifts observed in \cite{Eracleous_12} are the result of momentum imparted by a relativstic jet, the BLR clouds should have velocity offsets peculiar to the SMBH from which the broad lines originate.  If this  scenario is correct, the broad line peaks and Fe~K$\alpha$ line peaks should show different velocity shifts with respect to the host galaxy's rest frame.  However, in both the recoiling and binary SMBH scenarios, the Fe~K$\alpha$ line and the broad Balmer lines would show the same velocity  shifts with respect to the host galaxy's rest frame.  
These energy shifts should be detectable with \textit{Chandra's} High Energy Transmission Grating (HETG) X-ray spectrometer, which can achieve precision on Fe~K$\alpha$ line energies down to $\sim$~1~eV \citep[e.g.,][]{yaqoob+04,young+05,shu+10}, and the expected range of Doppler shifts for the \cite{Eracleous_12} sample is $\sim$~20~$-$~100~eV (corresponding to velocities of 1,000-5,000~$\mathrm{km\ s^{-1}}$).  
The following seven quasars from the \cite{Eracleous_12} sample are currently members of the \textit{Chandra} Source Catalog \citep[CSC 2.0,][]{evans+20}: J020011, J092712, J093844, J110556, J124551, J140251, and J151443.  However, the \textit{Chandra} exposure times for these quasars are relatively shallow; deep exposure times resulting in high X-ray photon counts are necessary to detect broad Fe~K$\alpha$ lines in AGN \citep[e.g.,][where the likelihood is raised to $>42$\% for sources with at least 10,000 counts in their 2-10~keV spectra]{guainazzi06}. 
 
Yet another test one can use to disciminate BLR outflows from SMBH dynamics is the reverberation of the optical Balmer lines.  If the blueshifted Balmer lines are produced in outflows directed towards the observer, they should respond to variations in the central engine's ionizing continuum with no time lag. Conversely, if the redshifted Balmer lines are produced in outflows directed away from the observer, they should respond to
variations in the ionizing continuum with a measureable time lag corresponding to the light travel time delay (i.e., the delay corresponding to the path length difference for light travelling directly to the observer from the accretion disk and light first travelling to excite Balmer transitions in gas flowing away from the observer).
\vspace{-0.5mm}

\pagebreak
\section{Summary \& Conclusions} \label{sec:conclusion}
In this paper we have presented the results from the VLBA X-band (8.5~GHz) observations of 34/88 of the \cite{Eracleous_12} binary SMBH candidates.  We detected 11 of the targets with the VLBA as unresolved, compact sources,  and six targets as (marginally) partially resolved sources (consistent with emission dominated by the radio core of an AGN).  These partially resolved sources will need future higher-resolution follow-up in order to better identify the nature of this emission that extends beyond the synthesized beamwidths (e.g., a double radio core, a core and a jet component, or potentially a young radio lobe).  These sources were originally identified as binary SMBH candidates in \cite{Eracleous_12} by virtue of the $\gtrsim1,000$~$\mathrm{km\ s^{-1}}$ offsets of the broad H$\beta$ emission lines with respect to the host galaxy's rest frame.  The main hypotheses for the velocity-offset broad lines are Doppler shifts induced by a relativistic jet interacting with BLR clouds, or the dynamics of a binary or recoiling SMBH.  While we find the former scenario unlikely, we defer presenting the bulk of the radio evidence against this model for the upcoming publication where we will present the results from the full VLA finder survey.  The true nature of these binary SMBH candidates has a major impact on the expected source population for the PTA and LISA gravitational-wave observatories, where recent physically motivated binary SMBH population modelling based upon this sample leads to a prediciton for the gravitational-wave background which is consistent with limits placed by PTAs \citep[][]{nguyen+20}.

We found three quasars with statistically significant, $\sim$10~pc radio/optical offsets which are interesting, but inconclusive. Future space-based observations with the HST (or JWST) would help to subtract the overwhelming contribution from quasar light and accurately model the host's optical/IR photometric center (due to the sufficiently well-characterized PSF and high-angular resolution possible with these facilities).  This type of analysis will allow us to better determine the origin of these offsets, and ascertain if  there are any significant radio/optical offsets hidden by the finite angular resolution and quasar contamination in the {\it Gaia}/SDSS data.  

While we do not directly find evidence for either the recoiling or binary  SMBH scenario in this study, we do not find evidence against such scenarios either.  Future follow-up with deep, higher angular resolution radio observations can help in this regard.  In particular, higher-frequency observations with the VLBA can improve on the angular resolution of the X-band data presented here, while also allowing for the increased  likelihood of detecting an inverted spectrum core under-luminous at X-band frequencies.  Similarly, the addition of data from the EVN for our current X-band data set, or ALMA (for VLBA data sets at frequencies from Q-band or above) would increase the maximum baseline of the visibilities (in comparison to the X-band data presented here) and achieve superior angular resolutions.  In this vein, the ngVLA or the EHT would also be premier observatories to test the binary SMBH nature of our candidates, capable of achieving spatial resolutions down to sub-pc scales for all of our sources (albeit with the EHT's very limited sensitivity).  Furthermore, X-ray observations aimed at the detection of Fe~K${\alpha}$ lines will help to distinguish between these different physical models, and gain insight into the physical properties of these systems.  Currently, \textit{Chandra's} HETG X-ray spectrometer is expected to achieve energy resolutions sufficient to detect the Doppler shifts expected for the \cite{Eracleous_12} sample.  However, NASA's planned \textit{Athena} X-ray spectrometer will improve on \textit{Chandra's} HETG energy resolution (by $\sim$ an order of magnitude) and sensitivity (at the relevant Fe~K${\alpha}$ line energies), and would be an ideal facility to measure Fe~K${\alpha}$ line profiles and energy shifts in the \cite{Eracleous_12} sample of binary SMBH candidates.

\acknowledgments

We thank the anonymous referee for a speedy review and for helpful comments and suggestions.  This work was done utilizing the publicly available data stets BS231 and VLA/12B-303, respectively collected by the VLBA and VLA instruments operated by the National Radio Astronomy Observatory (NRAO) (data is available at the following website: http://archive.nrao.edu/archive).  The NRAO is a facility of the National Science Foundation operated under cooperative agreement by Associated Universities, Inc.

This work has also made use of data from the European Space Agency (ESA) mission
{\it Gaia} (\url{https://www.cosmos.esa.int/gaia}), processed by the {\it Gaia}
Data Processing and Analysis Consortium (DPAC,
\url{https://www.cosmos.esa.int/web/gaia/dpac/consortium}). Funding for the DPAC
has been provided by national institutions, in particular the institutions
participating in the {\it Gaia} Multilateral Agreement. 

We acknowledge use of data from the Sloan Digital Sky Survey.  Funding for the Sloan Digital Sky Survey IV has been provided by the Alfred P. Sloan Foundation, the U.S. Department of Energy Office of Science, and the Participating Institutions. SDSS acknowledges support and resources from the Center for High-Performance Computing at the University of Utah. The SDSS web site is www.sdss.org.

M.E.\ and S.S.\ were supported by grant AST-1211756 from the National Science Foundation at the early stages of this work. T.B.\ acknowledges support by the National Science Foundation under Grant No. NSF AST-1211677  in the early stages of this research project. T.B. also acknowledges the support by the National Aeronautics and Space Administration (NASA) under award No.\ 80NSSC19K0319 and by the National Science Foundation (NSF) under award No.\ 1908042.
Part of this research was carried
out at the Jet Propulsion Laboratory, California Institute of
Technology, under a contract with the National Aeronautics and Space Administration.
The NANOGrav project receives support from National
Science Foundation (NSF) Physics Frontiers Center
award number 1430284.

\vspace*{5mm}

\facilities{VLBA, VLA, \textit{Gaia}, SDSS}
\software{CASA, AIPS}

\bibliography{bib}{}

\bibliographystyle{aasjournal}



\end{document}